\documentclass[11pt]{article}
\usepackage{sublabel}
\usepackage{latexsym}
\usepackage{amsthm}
\usepackage{amssymb}
\usepackage{graphicx}
\usepackage{amsmath}
\usepackage{pstricks}
\usepackage{fullpage}
\newpsobject{showgrid}{psgrid}{subgriddiv=1,griddots=10,gridlabels=6pt}

\newcommand {\abs} [1] {\left| #1 \right|}

\newcommand {\ie} {i.e.}

\numberwithin{equation}{section}

\title{The dynamics of transition to turbulence in plane Couette
flow}
\date{January 2007}
\author{D. Viswanath \thanks{Department of Mathematics,
University of Michigan, 
530 Church Street, Ann Arbor, MI 48109, U.S.A. }}

\begin{document}
\maketitle

\begin{abstract}
In plane Couette flow, the incompressible fluid between two plane
parallel walls is driven by the motion of those walls. The laminar
solution, in which the streamwise velocity varies linearly in the
wall-normal direction, is known to be linearly stable at all Reynolds
numbers ($Re$). Yet, in both experiments and computations, turbulence
is observed for $Re \gtrsim 360$. 

In this article, we show that for certain {\it threshold}
perturbations of the laminar flow, the flow approaches either steady
or traveling wave solutions. These solutions exhibit some aspects of
turbulence but are not fully turbulent even at $Re=4000$. However,
these solutions are linearly unstable and flows that evolve along
their unstable directions become fully turbulent.  The solution
approached by a threshold perturbation could depend upon the nature of
the perturbation. Surprisingly, the positive eigenvalue that
corresponds to one family of solutions decreases in magnitude with
increasing $Re$, with the rate of decrease given by $Re^{\alpha}$ with
$\alpha \approx -0.46$.
\end{abstract}

\newpage

\section{Introduction}
\subsection{Transition to turbulence}
The classical problem of transition to turbulence in fluids has not
been fully solved in spite of  attempts spread over more than a
century. Transition to turbulence manifests itself in a simple and
compelling way in experiments. For instance, in the pipe flow
experiment of Reynolds (see \cite{Acheson}), a dye injected at the
mouth of the pipe extended in ``a beautiful straight line through
the tube'' at low velocities or low Reynolds numbers ($Re$). The line
would shift about at higher velocities, and at yet higher velocities
the color band would mix up with the surrounding fluid all at once at
some point down the tube.

A wealth of evidence shows that the incompressible Navier-Stokes
equation gives a good description of fluid turbulence. Therefore one
ought to be able to understand the transition to turbulence using
solutions of the Navier-Stokes equation.  However, the nature of the
solutions of the Navier-Stokes equation is poorly understood. Thus the
problem of transition to turbulence is fascinating both
physically and mathematically.

The focus of this paper is on plane Couette flow. In plane Couette
flow, the fluid is driven by two plane parallel walls.  If the fluid
is driven hard enough, the flow becomes turbulent.  Such wall driven
turbulence occurs in many practical situations such as near the
surface of moving vehicles and is technologically important.

 The two parallel walls are assumed to be at $y=\pm 1$.  The walls move in
the $x$ or streamwise direction with velocities equal to $\pm 1$. The
$z$ direction is called the spanwise direction.  The Reynolds number
is a dimensionless constant obtained as $Re = UL/\nu$, where $U$
is half the difference of the wall velocities, $L$ is half the
separation between the walls, and $\nu$ is the viscosity of the
fluid. The velocity of the fluid is denoted by ${\bf u} = (u,v,w)$,
where $u,v,w$ are the streamwise, wall-normal, and spanwise
components.

 For the laminar solution, $v=w=0$ and $u=y$. The laminar solution is
linearly stable for all $Re$. As shown by Kreiss et al.\!\! \cite{KLH},
perturbations to the laminar solution that are bounded in amplitude
by $O(Re^{-21/4})$ decay back to the laminar solution. However, in
experiments and in computations, turbulent spots are observed around
$Re=360$ \cite{BTAA}.  The transition to turbulence in such
experiments must surely be because of the finite amplitude of the
disturbances. By a threshold disturbance, we refer to a disturbance
that would lead to transition if it were slightly amplified but which
would relaminarize if slightly attenuated. The concept of the
threshold for transition to turbulence was introduced by Trefethen 
and others \cite{TTRD}. The amplitude of the threshold disturbance depends
upon the type of the disturbance. It is believed to scale with
$Re$ at a rate given by $Re^\alpha$ for some $\alpha <= -1$.


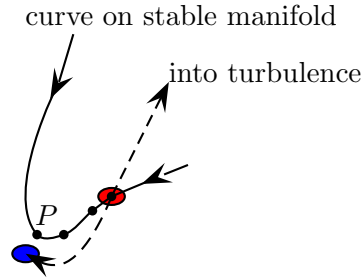
\begin{figure}[tb]
\begin{center}
\psset{xunit=0.5in,yunit=0.5in}
\begin{pspicture}(0.7,0.7)(4,4)
\psellipse[fillstyle=solid,fillcolor=blue](1.1,1.1)(0.15,0.1)
\psellipse[fillstyle=solid,fillcolor=red](2,1.7)(0.15,0.1)
\psline(1.6,3.4)(1.45,2.9)
\pscurve[arrowsize=8pt,showpoints=true]
{>-}%
(1.5,3)(1.22,1.3)(1.5,1.3)(1.8,1.55)(2,1.7)
\psline(2.8,2.03)(2.55,1.93)
\psline[arrowsize=8pt]
{>-}%
(2.6,1.95)(2,1.7)
\rput[bl](1.1,3.5){curve on stable manifold}
\psline[arrowsize=8pt,linestyle=dashed]
{->}%
(2,1.7)(2.6,2.9)
\pscurve[arrowsize=8pt,linestyle=dashed]
{->}%
(2,1.7)(1.5,1.0)(1.1,1.1)
\rput[bl](2.6,2.9){into turbulence}
\rput[bl](1.2,1.4){$P$}
\end{pspicture}
\end{center}
\caption[xyz]{
Schematic sketch of the dynamical picture of transition to turbulence
that is developed in this paper. The solid
oval stands for the laminar solution, and the empty oval stands for
a  steady or traveling wave solution.
}
\label{fig-1}
\end{figure}

Our main purpose is to explain how certain finite amplitude
disturbances of the laminar solution lead to turbulence.  The
dynamical picture that will be developed in this paper is illustrated
in Figure \ref{fig-1}. Historically, the laminar solution itself has
been the focus of attempts to understand mechanisms for
transition. Our focus however will be on a different solution that is
represented as an empty oval in Figure \ref{fig-1}.

Solutions that could correspond to the empty oval in Figure
\ref{fig-1} will be called lower-branch  solutions \cite{NagataA,
Waleffe2}. A solution at a certain value of $Re$ can be
continued by increasing a carefully chosen parameter. When this
parameter is increased, $Re$ first decreases and begins to increase
after a bifurcation point and we end up with an ``upper branch
solution'' at the original value of $Re$. The fact that a continuation
procedure can lead to an upper-branch solution appears to have no
significance for the dynamics at a fixed value of $Re$, however

 Depending upon the type of disturbance, the lower-branch
solution could either be a steady solution or a traveling wave.
Those solutions are not laminar in nature. Neither are they fully
turbulent even at high $Re$. Unlike the laminar solution, these
solutions are linearly unstable. The lower-branch 
solutions remain at an $O(1)$ distance from the laminar solution,
while the threshold amplitudes decrease with $Re$ as indicated
already. Therefore the threshold disturbances are too tiny to perturb
the laminar solution directly onto a lower-branch 
solution.  We will show, however, that some threshold disturbances
perturb the laminar solution to a point on the stable manifold of a
lower-branch  solution (point $P$ in Figure
\ref{fig-1}). A slightly larger disturbance brings the flow close to
the lower-branch  solution, after which the flow follows a
branch of its unstable manifold and becomes fully turbulent.

For certain types of  disturbances, the perturbed laminar solution
does not approach a lower branch solution. 
Thus the dynamical picture of Figure \ref{fig-1} is not valid for those
disturbances.
Instead it flows towards
an {\it edge state} \cite{SEY}. We give a brief discussion of the nature of the edge
states in Section 4.  

\subsection{Connections to earlier research}
The dynamical picture presented in Figure \ref{fig-1} is related
directly and indirectly to much earlier research. Basic results
from hydrodynamic stability show that some eigenmodes that
correspond to the least stable eigenvalue of the linearization around
the laminar solution do not depend upon the spanwise or $z$ direction.
This may lead one to expect that disturbances that trigger transition
to turbulence are $2$-dimensional. That expectation is not correct,
however. As shown by Orszag and Kells \cite{OK}, spanwise variation
is an essential feature of disturbances that trigger transition to
turbulence. Accordingly, all the disturbances considered in this
paper are $3$-dimensional.

Kreiss et al.\!\! \cite{KLH} and Lundbladh et al.\!\! \cite{LHR}
investigated disturbances that are non-normal pseudomodes of the
linearization of the laminar solution.  Since the laminar solution is
linearly stable, a slight perturbation along an eigenmode will simply
decay back to the laminar solution at a predictable rate.  The
pseudomodes are chosen to maximize transient growth of the solution of
the linearized equation, which is a consequence of the non-normality
of the linearization. Such disturbances lead to transition with quite
small amplitudes and will be considered again in this paper. It must
be noted, however, that any consideration based on the linearization
alone can only be valid in a small region around the laminar
solution. The dynamics of transition to turbulence, as sketched in
Figure \ref{fig-1}, involves an approach towards a lower-branch
 solution that lies at an $O(1)$ distance from the laminar
solution.  It is therefore necessary to work with the fully nonlinear
Navier-Stokes equation to explicate the dynamics of transition to
turbulence.

\begin{figure}
\begin{center}
\includegraphics[height=2.5in, width=3in]{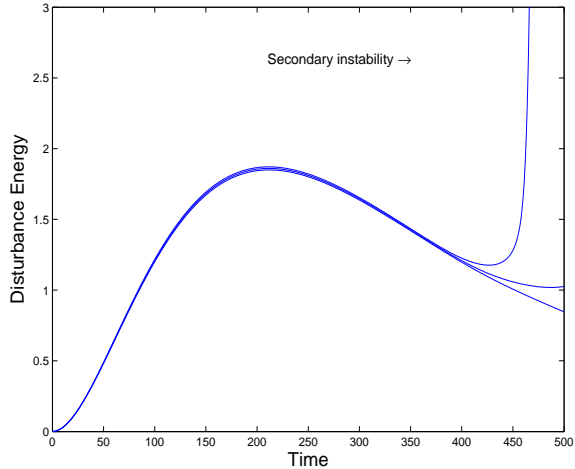}
\end{center}
\caption[xyz]{The plot above shows the secondary instability 
in a transition computation at $Re=2000$.}
\label{fig-2}
\end{figure}

Figure \ref{fig-2} shows the variation of the disturbance energy with
time for a disturbance that leads to transition.  We observe that the
disturbance energy increases smoothly initially and is then followed
by a spike. The spike is in turn followed by turbulence. The spike
corresponds to a secondary instability, as noted by Kreiss et al.\!\!
\cite{KLH}. In fact, the so-called secondary instability is just the
linear instability of a lower-branch  solution as will
become clear.

Partly motivated by the secondary instability, there was a 
search for nonlinear steady solutions related to transition as reviewed
in \cite{CE}. Early success in this effort was due to Nagata
\cite{NagataA, NagataB} who computed steady solutions of plane Couette
flow in the interval $125\leq Re \leq 300$.
Waleffe \cite{Waleffe1, Waleffe2, Waleffe3} introduced a more flexible
method for computing such solutions, and like Nagata, 
argued that such solutions
could be related to transition to turbulence. The numerical method
we use was introduced in \cite{Viswanath}. It uses a combination of
Krylov space methods and the locally optimally constrained hook step
to achieve far better resolution as show by \cite{GHC}, \cite{Viswanath},
\cite{WGW},
and this paper.

The computations in \cite{KLH, LHR} imply that threshold amplitudes
scale as $Re^\alpha$ for $\alpha < -1$. The value of $\alpha$ appears
to depend upon the type of perturbation. Our focus is not on
determining the scaling of the threshold amplitudes. Nevertheless, we
will discuss numerical difficulties that beset determination of
threshold amplitudes.

Measuring threshold amplitudes poses experimental challenges as well
and it is not always clear from experiments if the thresholds have 
a simple power scaling with $Re$. One difficulty is that the
turbulent states can be short lived. Schmiegel and Eckhardt \cite{SE}
have connected the lifetime of turbulence to the possibility that 
turbulent dynamics in the transition regime is characterized by 
a chaotic saddle and not a chaotic attractor.

\subsection{Connections to recent research}
Wang et al.\!\! \cite{WGW} have taken steps towards an asymptotic theory
of the lower branch solutions and carried their computation beyond
$Re=50,000$. 
They connect the asymptotics to scalings of the threshold for transition
to turbulence.
The lower branch states occur as
solutions to equations that use periodic boundary conditions. 
Because such boundary conditions cannot be realized in  laboratory
setups, the solutions are best thought of as waves. Thus it is
pertinent to consider their stability with respect to subharmonic
disturbances as in \cite{WGW}. That paper also suggests that 
lower branch solutions might be of use for control. A somewhat
different suggestion related to control can be found in \cite{Kawahara}.

Not all disturbances follow the dynamical picture of Figure
\ref{fig-1} as already noted.
For the third type of disturbance considered in Section 4, the laminar
solution perturbed by the threshold disturbance evolves towards a
state that looks almost like an invariant object of the underlying
differential equation. Those objects have been termed edge states by
Schnieder et al.\!\!
\cite{SEY}.  Lagha et al.\!\! \cite{LSLE} make the important
point that the dynamical picture of Figure \ref{fig-1} can be valid
for typical disturbances only if the lower-branch solution has a
single unstable eigenvalue.


 Near the threshold for the third type of disturbance,
it appears as if the disturbed state evolves and approaches a
traveling wave. Indeed, a crude or under-resolved computation could
easily mistake that appearance for a true solution.  When we attempted
to refine that near-solution using the numerical method reviewed in
Section 3, the numerical method converged to a traveling wave
solution. However, that traveling wave has two unstable eigenvalues
and the flow near the threshold does not come as close to that
traveling wave as the dynamical picture of Figure \ref{fig-1} would
require. The disturbed state appears to evolve into an edge state.

Visualizing the dynamics in state space is fundamental to the approach
to transition to turbulence sketched in this paper and in the articles
discussed above. Yet there has so far been no way to obtain revealing
visualizations of state space dynamics.    Gibson et al.\!\!
\cite{GHC} have recently produced revealing visualizations
of the state space of turbulent flows. For instance, one of their
figures shows a messy-looking  turbulent trajectory cleanly trapped
by the unstable manifolds of certain equilibrium solutions. Such
visualizations might prove useful to both computational and
experimental investigations of transition to turbulence

Section 2 reviews some basic aspects of plane Couette flow. The
numerical method used to flesh out the dynamical picture of Figure
\ref{fig-1} is given in Section 3. In Section 4, we consider three different types of disturbances.  The
lower-branch solutions (empty oval of Figure \ref{fig-1})
that correspond to the first two types are steady solutions. For a
given $Re$, the solutions that correspond to these two types are
identical modulo certain symmetries of plane Couette flow.
In Section 5, we consider some qualitative aspects of the solutions
reported in Section 4.  A surprising finding is that these these
solutions are less unstable for larger $Re$. The top eigenvalue of
these solutions is real and positive. For one family of solutions, the
top eigenvalue appears to decrease at the rate $Re^\alpha$ for $\alpha
\approx -0.46$.

In the concluding Section 6, we give additional context for this paper
from two points of view. The first point of view is mainly
computational and has to do with reduced dimension methods.  In this
paper, we have taken care to use adequate spatial resolution to ensure
that the computed solutions are true solutions of the Navier-Stokes
equation. We recognize, however, that resolving all scales may prove
computationally infeasible in some practical situations. We argue that
transition to turbulence computations can be useful in gaging the
possibilities and limitations of methods that do not resolve all
scales.  Secondly, we briefly discuss the connection of transition
computations with transition experiments.

\section{Some aspects of plane Couette flow}
The Navier-Stokes equation 
$\partial {\bf u}/\partial t + ({\bf u}.\nabla){\bf u} =
-(1/\rho)\nabla p + (1/Re)\triangle {\bf u}$
describes the motion of incompressible fluids. The velocity
field ${\bf u}$ satisfies the incompressible constraint
$\nabla.{\bf u}=0$. For plane Couette flow the boundary 
conditions are ${\bf u} = (\pm 1, 0, 0)$ at the walls,
which are at $y=\pm 1$. To render the computational domain
finite, we impose periodic boundary conditions in the $x$
and $z$ directions, with periods $2\pi \Lambda_x$ and
$2\pi \Lambda_z$, respectively. To enable comparison with
\cite{LHR}, we use $\Lambda_x = 1.0$ and $\Lambda_z=0.5$
throughout this paper. 

Certain basic quantities are useful for forming a general idea of the
nature of a velocity field of plane Couette flow.  The first of these
is the rate of energy dissipation per unit volume for plane Couette
flow, which is given by
\begin{equation}
D = \frac{1}{8\pi^2 \Lambda_x\Lambda_z}\int_0^{2\pi\Lambda_z}
\int_{-1}^{+1} \int_0^{2\pi\Lambda_x} \abs{\nabla u}^2
+\abs{\nabla v}^2 + \abs{\nabla w}^2 \: dx\, dy\, dz.
\label{eqn-2-1}
\end{equation}
The rate of energy input per unit volume is given by
\begin{equation}
I = \frac{1}{8\pi^2 \Lambda_x\Lambda_z}\int_0^{2\pi\Lambda_x}
\int_0^{2\pi\Lambda_z} \frac{\partial u}{\partial y} \Bigl\lvert_{y=1}
+ \frac{\partial u}{\partial y} \Bigl\lvert_{y=-1} \: dx\, dz.
\label{eqn-2-2}
\end{equation}
For the laminar solution $(u,v,w) = (y,0,0)$, both $D$ and $I$ are
normalized to evaluate to $1$.  Expressions such as \eqref{eqn-2-1}
and \eqref{eqn-2-2} are derived using formal manipulations. The
derivations would be mathematically valid if the velocity field ${\bf
u}$ were assumed to be sufficiently smooth. Although such smoothness
properties of solutions of the Navier-Stokes  are yet to be proved,
numerical solutions possess the requisite smoothness. Even
solutions in the turbulent regime appear to be real analytic in the
time and space variables, which is why spectral methods have been
so successful in turbulence computations.

 In the long run, on physical grounds, we expect the time averages of $D$
and $I$ to be equal because the energy dissipated through viscosity
must be input at the walls. For steady solutions and traveling waves,
the values of $D$ and $I$ must be equal.

Another useful quantity is the disturbance energy. The disturbance
energy of $(u,v,w)$ is obtained by integrating $(u-y)^2+v^2+w^2$
over the computational box. This quantity has already been used
in Figure \ref{fig-2}. The disturbance energy is a measure of the
distance from the laminar solution.

Two discrete symmetries of the Navier-Stokes equation for plane
Couette flow will enter the discussion later. The shift-reflection
transformation of the velocity field is given by
\begin{equation}
S_1{\bf u} = \begin{pmatrix}
u\\v\\-w
\end{pmatrix}
\Biggl(x+\pi \Lambda_x, y, -z\Biggr),
\label{eqn-2-3}
\end{equation}
and the shift-rotation transformation of the velocity field is
given by
\begin{equation}
S_2{\bf u} = \begin{pmatrix}
-u\\-v\\w
\end{pmatrix}
\Biggl(-x+\pi \Lambda_x, -y, z + \pi \Lambda_z\Biggr).
\label{eqn-2-4}
\end{equation}
Plane Couette flow is unchanged under both these transformations.
Thus if a single velocity field along a trajectory of plane Couette
flow satisfies either symmetry, all points along the trajectory must
have the same symmetry. However, velocity fields that lie on
the stable and unstable manifolds of symmetric periodic or relative
periodic solutions need not be symmetric.

\section{Numerical method}
The Navier-Stokes equation in the standard form given in Section 2
cannot be viewed as a dynamical system because the velocity
field ${\bf u}$ must satisfy the incompressibility condition
and because there is no equation for evolving the pressure
$p$. It can be recast as a dynamical system, however, by
using the $y$ components of ${\bf u}$ and $\nabla \times {\bf u}$,
which is the vorticity field. If the resulting system is discretized
in space using $M+1$ Chebyshev points in the $y$ direction,
and $2L$ and $2N$ Fourier points in the $x$ and $z$ directions, respectively,
the number of degrees of freedom of the spatially discretized system
is given by 
\begin{equation}
2(M-1)+(2M-4)((2N-1)(2L-1)-1)
\label{eqn-3-1}
\end{equation}
as shown in \cite{Viswanath}. We do not use a truncation strategy
to discard  modes and we employ dealiasing in the directions
parallel to the wall.

\begin{figure}
\begin{center}
\includegraphics[height=2.5in, width=3in]{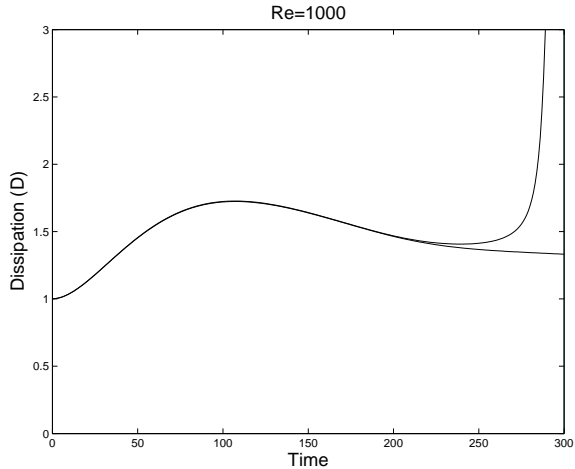}
\end{center}
\caption[xyz]{The plot above shows the variation of $D$ defined
by \eqref{eqn-2-1} for a disturbance slightly above the
threshold and for a disturbance slightly below the threshold.}
\label{fig-3}
\end{figure}

Given a form of the disturbance $P$, the threshold for
transition is obtained by integrating the disturbed
velocity $(y,0,0)+\epsilon P$ in time for different $\epsilon$
\cite{KLH}. If $\epsilon$ is greater than the threshold value,
the flow will spike and become turbulent as evident from
Figures \ref{fig-2} and \ref{fig-3}. If $\epsilon$ is below
the threshold value, the flow will relaminarize. 
As indicated by Figures \ref{fig-2} and \ref{fig-3},
we may graph either disturbance energy or $D$ to 
examine a value of $\epsilon$. We may also graph
$I$, which is defined by \eqref{eqn-2-2}, against time.

The accurate determination of thresholds is beset by numerical
difficulties. To begin with, suppose that we are able to integrate the
Navier-Stokes equation for plane Couette flow exactly. Then as implied
by the dynamical picture in Figure \ref{fig-1}, a disturbance of the
laminar solution that is on the threshold will fall into a
lower-branch  solution, and it will take infinite time to do
so. However, computations for determining the threshold, such as that
shown in Figure \ref{fig-2}, can only be over a finite interval of
time. Thus the finiteness of the time of integration is a source of
error in determining thresholds. Two other sources of error are
spatial discretization and time discretization.

An accurate determination of the threshold will need to estimate and
balance these three sources of error carefully. In our computations,
we determine the thresholds with only about $2$ digits of accuracy.
That modest level of accuracy is sufficient for our purposes. In
Tables \ref{table-1} and \ref{table-3}, the thresholds are reported
using disturbance energy per unit volume.

Once the threshold has been determined, we need to compute a steady
solution or a traveling wave to complete the dynamical picture of
Figure \ref{fig-1}. The initial guess for that lower-branch 
solution is produced by perturbing the laminar solution by adding the
numerically determined threshold disturbance and integrating the
perturbed point over the time interval used for determining the
threshold (this time interval is $500$ in Figure \ref{fig-2} and $300$
in Figure \ref{fig-3}).

That initial guess is fed into the method described in
\cite{Viswanath} to find a lower-branch  solution with good
numerical accuracy. That method finds solutions by solving Newton's
equations, but the equations are set up and solved in
a non-standard way. Suppose
that the spatially discretized equation for plane Couette flow is
written as $\dot{x} = f(x)$, where the dimension of $x$ is given by
\eqref{eqn-3-1}. To find a steady solution, for instance,
it is natural to solve $f(x)=0$ after supplementing that equation by
some conditions that correspond to the symmetries \eqref{eqn-2-3} and
\eqref{eqn-2-4}. However that is not the way we proceed.  We solve for
a fixed point of the time $t$ map $x(t; x_0)$, for a fixed value of
$t$, after accounting for the symmetries. The Newton equations are
solved using GMRES. The method does not always compute the full Newton
step, however. Instead, the method finds the ideal trust region step
within a Krylov subspace as described in \cite{Viswanath}.

This method can easily handle more than $10^5$ degrees of freedom, and
thus makes it possible to carry out calculations with good spatial
resolution. The reason for setting up the Newton equations in the
peculiar way described in the previous paragraph has to do with the
convergence properties of GMRES. The matrix that arises in solving the
Newton equations approximately has the form $I - \partial x(t;x_0)/
\partial x_0$, where $I$ is the identity. Because of viscous damping
of high wavenumbers, many of the eigenvalues of that matrix will be
close to $1$, thus facilitating convergence of GMRES. We may expect
the convergence to deteriorate as $Re$ increases, because viscous
damping of high wavenumbers is no longer so pronounced, and that is
indeed the case. Nevertheless, we were able to go up to $Re=4000$, and
we believe that even higher values of $Re$ can be reached.

\section{Disturbances of the laminar solution and transition to
turbulence}

In this section, we consider three types of disturbances and determine
the threshold amplitudes for various values of $Re$. To complete the
dynamical picture of Figure \ref{fig-1}, we determine for the
first two types
the steady solution
or traveling wave that corresponds to the empty oval of that figure
using the numerical method of the previous section. 

\subsection{Rolls with unsymmetric noise}

\begin{table}
\begin{center}
\begin{tabular}{c|c|c|c|c|c|c}
Label & $Re$ & $D$/$I$ & $\lambda_{max}$ & $Re_\tau$ & $T$ & threshold\\ \hline
$B1$& $500$ & $1.3920$  & $.04326$ & $53$ & $150$   & $2.46e-4$ \\ \hline
$B2$& $1000$ & $1.3486$ & $.03294$ & $73$ & $300$   & $5.73e-5$ \\ \hline
$B3$& $2000$ & $1.3285$ & $.02413$ & $103$ & $500$  & $1.36e-5$ \\ \hline
$B4$& $4000$ & $1.3210$ & $.01732$ & $145$ & $1000$ & $3.30e-6$ 
\end{tabular}
\end{center}
\caption[xyz]{Data for disturbances of the form \eqref{eqn-4-1} with
unsymmetric noise and  for steady solutions that correspond to the empty
oval in Figure \ref{fig-1}. The steady solutions
are labeled $B1$ through $B4$. 
$D$ and $I$, which are defined by \eqref{eqn-2-1} and \eqref{eqn-2-2},
correspond to those steady solutions. The next two columns give
the eigenvalue with the maximum real part and the frictional Reynolds
number for those solutions. $T$ is the time interval used to determine
the threshold disturbance and the threshold is reported using
disturbance energy per unit volume.
}
\label{table-1}
\end{table}

We follow \cite{KLH} and consider the disturbance,
\begin{equation}
(u, v, w) = \epsilon (0, \psi_z, -\psi_y),
\label{eqn-4-1}
\end{equation}
where $\psi = (1-y^2)^2 \sin(z/\Lambda_z)$. This disturbance is
unchanged by both $S_1$, which was defined by \eqref{eqn-2-3},
and by $S_2$, which was defined by \eqref{eqn-2-4}.
A disturbance of the laminar solution ${\bf u} = (y, 0, 0)$ of
the form \eqref{eqn-4-1} never leads to transition to turbulence.
It is necessary to add some more terms to the disturbance to 
make the velocity field depend upon the $x$ direction.

To introduce dependence on $x$, we add modes of the Stokes problem.
One can get an eigenvalue problem for $\hat{v}(y)$, where
$v = \hat{v}(y) \exp(\iota l x/\Lambda_x + \iota n z/\Lambda_z)
\exp(\sigma t)$, or for $\hat{\eta}(y)$, where
$\eta = \hat{\eta}(y) \exp(\iota l x/\Lambda_x + \iota n z/\Lambda_z)
\exp(\sigma t)$. Here $\eta$ is the wall-normal component of the
vorticity field.  For a $v$ mode, $\eta=0$, and vice versa.  For a
given mode, the velocity field is recovered using the
divergence free condition.   The velocity fields of modes
with different $(l,n)$ are obviously orthogonal. A calculation shows
that the velocity fields for the $v$ and $\eta$ modes with the same
$(l,n)$ are also orthogonal. For a given $(l,n)$, we pick the $v$ and
$\eta$ modes with the least stable $\sigma$.

To the disturbance \eqref{eqn-4-1},
we added both $v$ and $\eta$ modes for $(l,n)$ with $-3\leq l \leq 3$
and $-7 \leq n \leq 7$. Together the added modes can be called noise.
The energy of the noise was equal to $1\%$ of the energy of
\eqref{eqn-4-1}. This energy was equally distributed over the various
orthogonal modes. Following \cite{KLH}, we chose random phases for
the modes. The threshold can depend upon the choice of phase. 
Therefore, for accurate determination of thresholds it is better
to use non-random phases.

After adding modes of this form to \eqref{eqn-4-1}, the resulting
disturbance in unchanged by neither $S_1$ nor $S_2$. Therefore the
disturbance is unsymmetric. Table \ref{table-1} reports data from
computations carried out using such an unsymmetric disturbance.  The
thresholds in that table give the energy of \eqref{eqn-4-1} and do not
include the energy within the noise terms.  The lower-branch
 solutions $B1$ through $B4$ correspond to the empty oval in
Figure \ref{fig-1}. Each of these solutions appears to have a single
unstable eigenvalue. We determined the most unstable eigenvalues using
simultaneous iteration and the time $t$ map of the Navier-Stokes
equation, as in Section 3, with $t=8$. All the solutions seem to have
just one unstable eigenvalue. That eigenvalue is real. Surprisingly,
it decreases with $Re$ at the rate $Re^\alpha$, where $\alpha \approx
-0.46$.  Thus the lower-branch  solutions become less and
less unstable with increasing $Re$.

\begin{figure}
\begin{center}
\includegraphics[height=2.1in,width=2.1in]{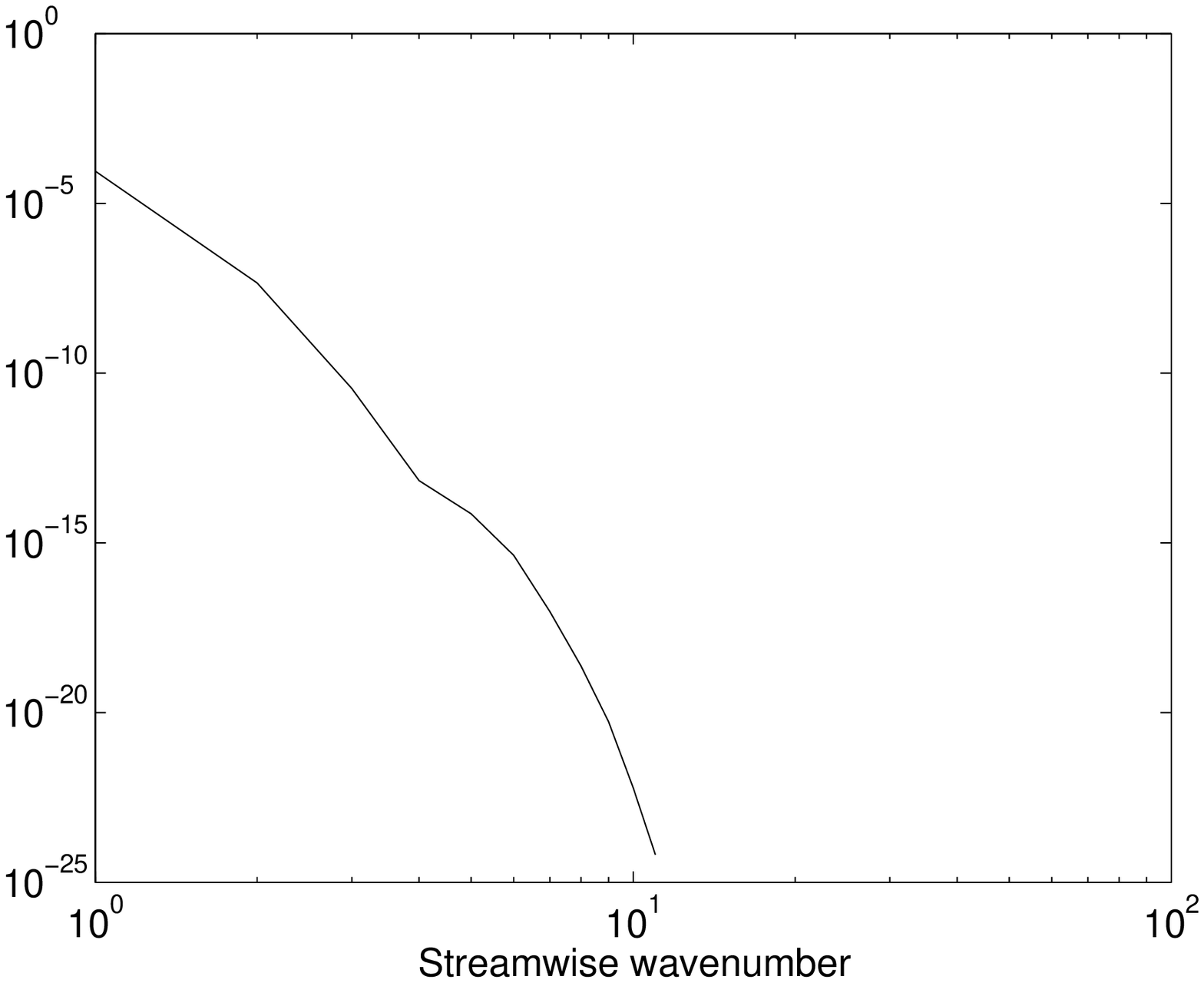}
\includegraphics[height=2.1in,width=2.1in]{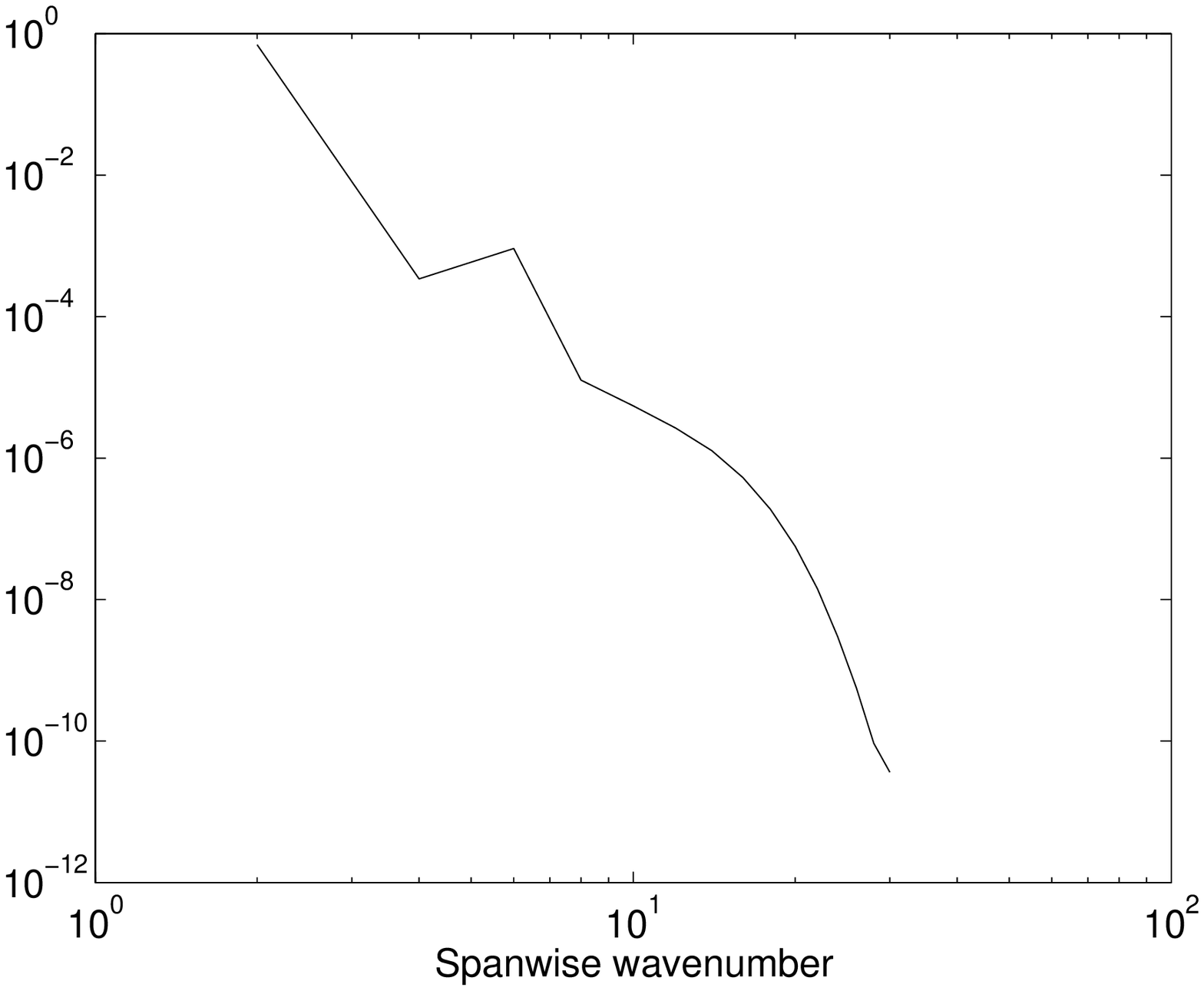}
\includegraphics[height=2.1in,width=2.1in]{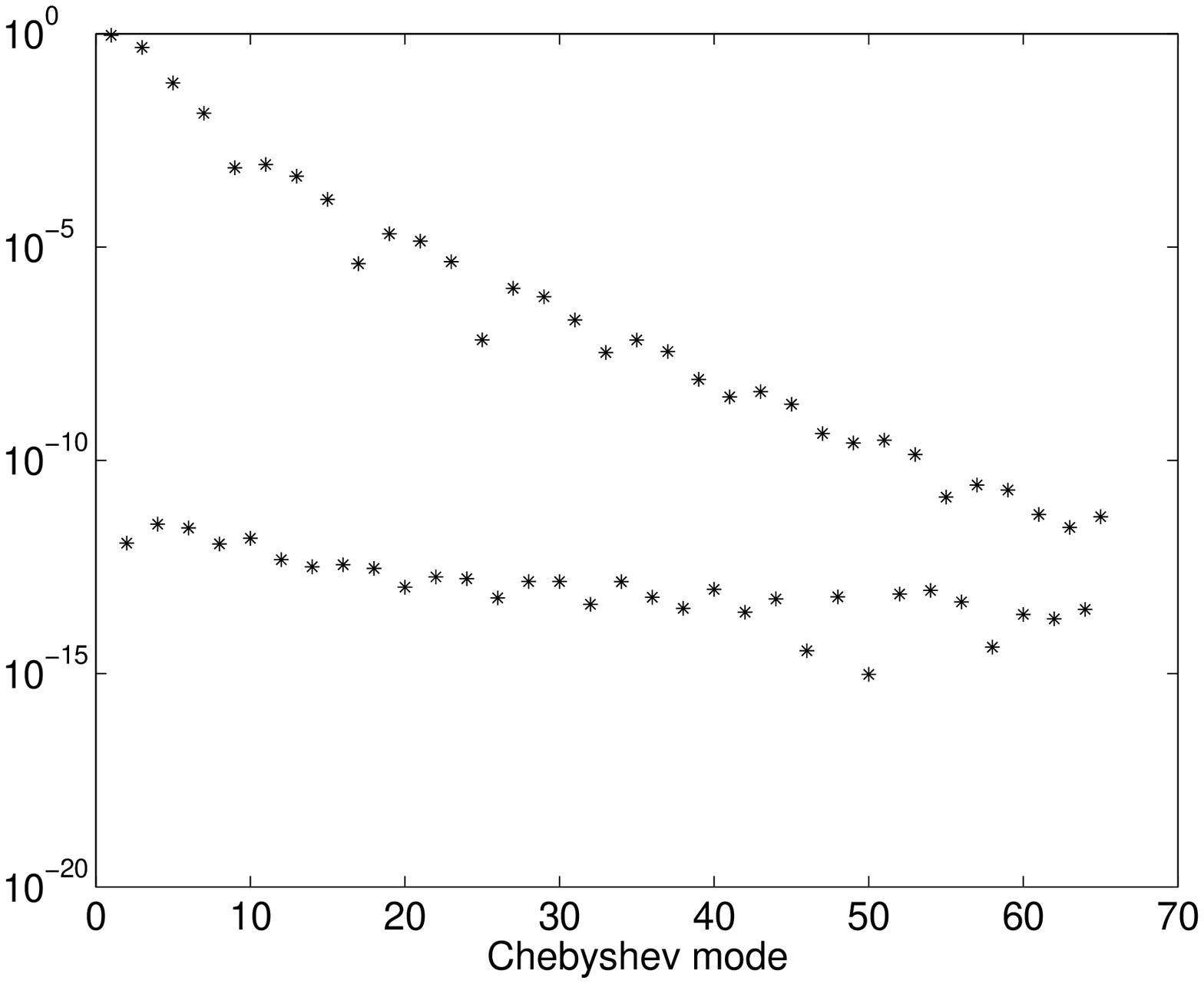}
\end{center}
\caption[xyz]{The plots above graph the energy in the solution
$B4$ of Table \ref{table-1} against streamwise wavenumber,
spanwise wavenumber, and Chebyshev mode.}
\label{fig-4}
\end{figure}

All our computations used $(2L, M+1, 2N) = (24, 65, 32)$.  By
\eqref{eqn-3-1}, the number of degrees of freedom in the computation
for finding the lower-branch  solutions is $88414$. As shown
by Figure \ref{fig-4}, that much resolution was entirely adequate. The
solutions $B1$ through $B4$ were computed with at least $5$ digits of
accuracy.

\subsection{Rolls with symmetric noise}

\begin{table}
\begin{center}
\begin{tabular}{c|c|c|c|c}
Label & $Re$ & $s_x$ & $s_z$ & Threshold \\ \hline
$C1$ & $500$  & $1.5600$ & $.0016$ & $2.97e-4$ \\ \hline
$C2$ & $1000$ & $6.1093$ & $.0012$ & $5.72e-5$\\ \hline
$C3$ & $2000$ & $0.5075$ & $.0018$ & $1.40e-5$\\ \hline
$C4$ & $4000$ & $2.8719$ & $.0013$ & $3.28e-6$
\end{tabular}
\end{center}
\caption[xyz]{Data for disturbances of the form \eqref{eqn-4-1} with
symmetric noise and  steady solutions that correspond to the empty
oval in Figure \ref{fig-1}.
The solutions $Cn$ are connected to the solutions $Bn$ of 
Table \ref{table-1} as follows: $Cn(x+s_x, y, z + s_z) = Bn$.
}
\label{table-2}
\end{table}

It has been suggested that one purpose of adding the noise to
\eqref{eqn-4-1} is to break symmetries and that a symmetric
disturbance would lead to drastically increased thresholds
\cite{KLH}. To investigate that matter, we symmetrized the
disturbances used to generate Table \ref{table-1}. More specifically,
if ${\bf u}$ is a disturbed velocity field, we replaced it by $({\bf
u} + S_1 {\bf u} + S_2 {\bf u} + S_1 S_2 {\bf u})/4$ which is unchanged
by both $S_1$ and $S_2$.  A comparison of Tables \ref{table-1} and
\ref{table-2} shows that the thresholds are in fact not elevated.
Thus we conclude that the purpose of adding the noise is not to break
the symmetry but to introduce dependence on the $x$ direction.  The
lower-branch  solutions that correspond to such symmetric
disturbances are labeled $C1$ through $C4$ in Table \ref{table-2}.

The solutions $C1$ though $C4$ are just translations of the
solutions $B1$ through $B4$ as indicated in Table \ref{table-2}. If
the thresholds were determined exactly, the disturbances of Tables
\ref{table-1} and \ref{table-2} would come arbitrarily close to the
corresponding solution in the infinite time limit. Each threshold in
those tables was determined inexactly using a finite time interval,
and we verified that the disturbed states evolve and come within $2\%$
of the corresponding lower-branch  solution. Thus there can
be little doubt about the role of these lower-branch 
solutions in the transition to turbulence. The $C$ family of solutions
is the same as the lower-branch family of \cite{Waleffe3}.

Given that the solutions $C1$ through $C4$ are just translations of
the solutions $B1$ through $B4$, it is tempting to think that all
threshold disturbances, say at $Re=4000$, might evolve and approach a
translate of a single solution such as $C4$. That is not correct,
however, as we will now show.

\subsection{Superposed Orr-Sommerfeld modes}

\begin{table}
\begin{center}
\begin{tabular}{c|c|c|c|c|c|c|c|c}
Label & $Re$ & $D/I$ & $\lambda_{max}$ & $Re_\tau$ & $c_x$ & $c_z$ & $T$ & Threshold \\ \hline
$D1$ & $500$ & $1.2863$  & $.0464$ & $51$ & $.3051$ & $0$ & $100$ & $8.4e-3$
\\ \hline
$D2$ & $1000$ & $1.2522$ & $.0379$ & $72$ & $.2666$ & $0$ & $200$ & $1.6e-3$
\end{tabular}
\end{center}
\caption[xyz]{Data for disturbances obtained by superposing 
Orr-Sommerfeld modes and for the corresponding traveling waves
labeled $D1$ and $D2$.
$c_x$ and $c_z$ give the wave speeds in the $x$ and $z$ directions.
The other columns are as in Table \ref{table-1}.
}
\label{table-3}
\end{table}

The disturbances for Table \ref{table-3} were obtained by superposing
Orr-Sommerfeld modes as in \cite{OK}. An Orr-Sommerfeld mode is
of the form $(u, v , w) = (\hat{u}(y), \hat{v}(y), \hat{w}(y))
\exp(\iota lx/\Lambda_x + \iota nz/\Lambda_z) \exp(\sigma t)$.
We use Orr-Sommerfeld modes with $(l,n) = \pm (1,0)$ and $(l, n) = \pm
(1,1)$.  The phases of the Orr-Sommerfeld modes were chosen to make
$\hat{v}(0)$ real.  The disturbance energy was equally distributed
across the modes. For given $(l,n)$, we chose the least stable mode
and symmetrized it as in Equation (3.2) of \cite{OK}. Note that the
disturbance depends on both the $x$ and $z$ directions.

The solutions obtained by following the numerical method of Section 3
were traveling waves in this case. The wave speeds for both $D1$ and
$D2$ in Table \ref{table-3} are nonzero in the $x$ direction. These
traveling waves are unsymmetric and they do not become symmetric even
after translations in the $x$ and $z$ directions.

The thresholds for this third type of disturbance are reported in
Table \ref{table-3}. Close to the threshold, the flow appears to
approach a traveling wave. After a diligent computation, we feel sure
that there is no true traveling wave solution or relative periodic
solution to complete the dynamical picture of Figure \ref{fig-1}. The
flow near the threshold evolves and comes within $10\%$ of $D1$ or
$D2$ but no closer. It appears to approach an edge state.

\begin{figure}
\begin{center}
\includegraphics[height=2.5in, width=3in]{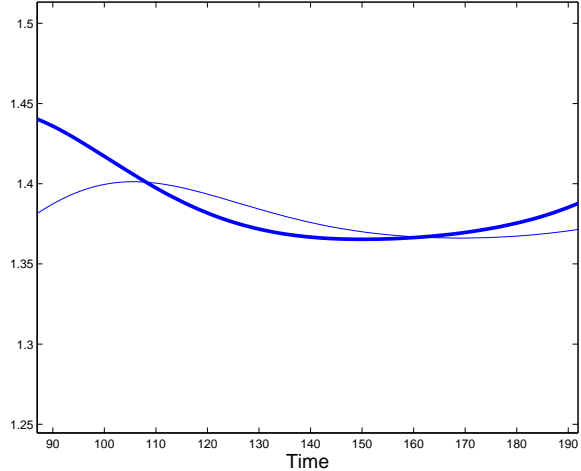}
\end{center}
\caption[xyz]{The thick line is a plot of $D$ defined by \eqref{eqn-2-1} and
the thin line is a plot of $I$ defined by \eqref{eqn-2-2}. The disturbance
used to get the plots was a superposition of Orr-Sommerfeld modes
at $Re=500$.}
\label{fig-x}
\end{figure}

Figure \ref{fig-x} shows plots of the rates of energy input and energy
dissipation near an edge state. In that figure, the disturbance is
very close to the threshold and the time axis is chosen to correspond
to an edge state. Note that the dissipation sags below energy input and
then rises above it. Therefore, we do not expect a traveling wave 
or an equilibrium solution near the edge state. The second
crossing of the two curves is below the first. In addition, both the
curves spike and transition to turbulence soon after they cross.
Therefore, a periodic or relative periodic solution is unlikely to be
found near the edge state.

In pipe flow transition computations, we have observed that the rate of
dissipation and the rate of energy input become almost horizontal lines
near the edge states. The rate of dissipation is slightly above 
the rate of energy input, suggesting that there may be no invariant objects
near the edge states in these instances.

As stated earlier, the laminar solution of plane Couette flow is
linearly stable. The computations of this section shed some light
on the laminar-turbulent separatrix. A part of this separatrix is
formed by the stable manifolds of the $B$ and $C$ family of solutions.
We have shown that these stable manifolds come closer and closer
to the laminar solution as $Re$ increases. The traveling waves
$D1$ and $D2$ are also on the separatrix. However, we have not 
found tiny disturbances to the laminar solution for which the thresholds
diminish in magnitude with increasing $Re$ and which approach these
solutions as the flow evolves as in Figure \ref{fig-1}. In the next section,
we show that the $D$ solutions are qualitatively similar to the $B$ and
$C$ solutions.

\section{Lower-branch  solutions of plane Couette flow}

\begin{figure}
\begin{center}
\includegraphics[height=1.8in, width=2.1in]{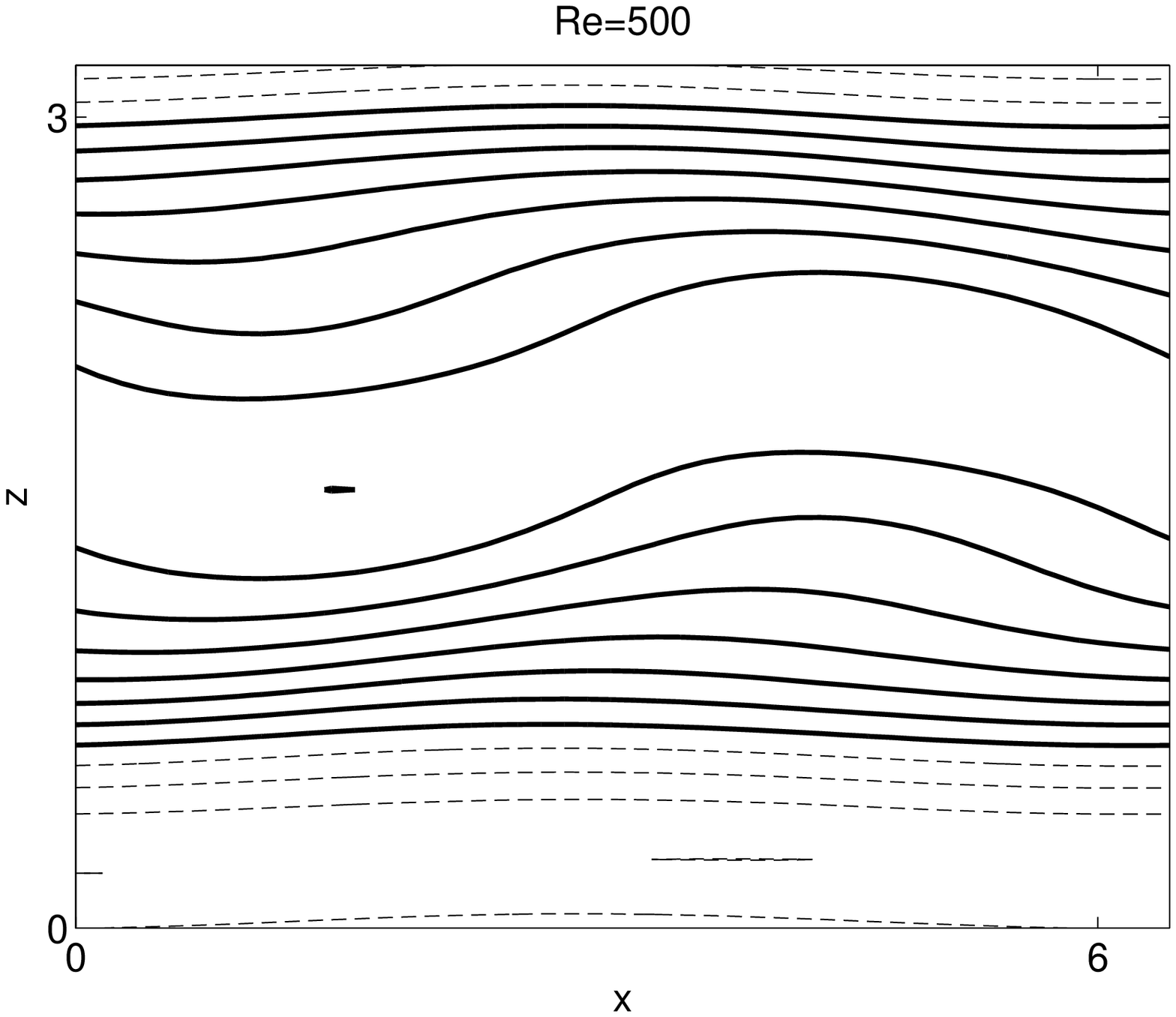}
\includegraphics[height=1.8in, width=2.1in]{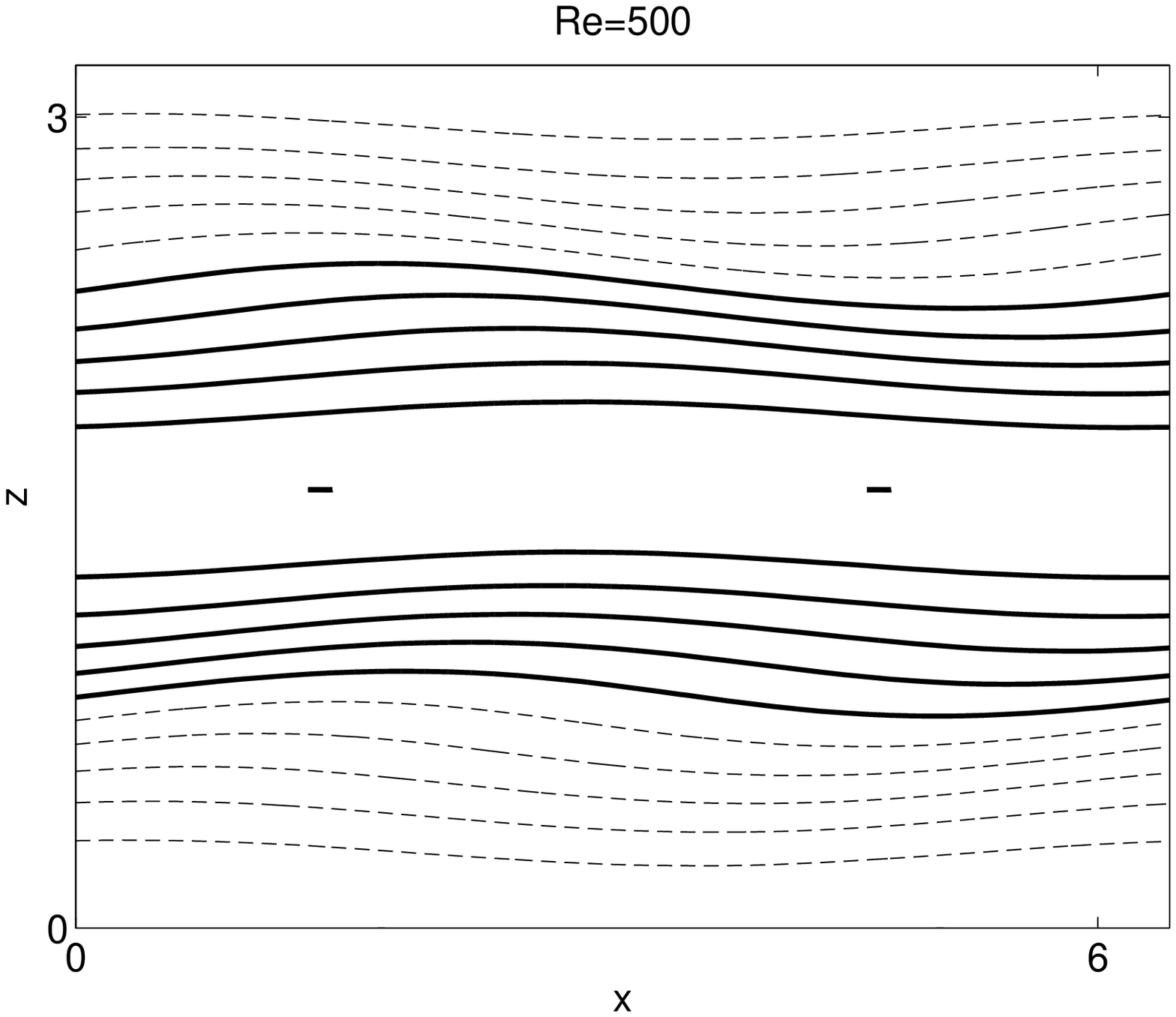}
\includegraphics[height=1.8in, width=2.1in]{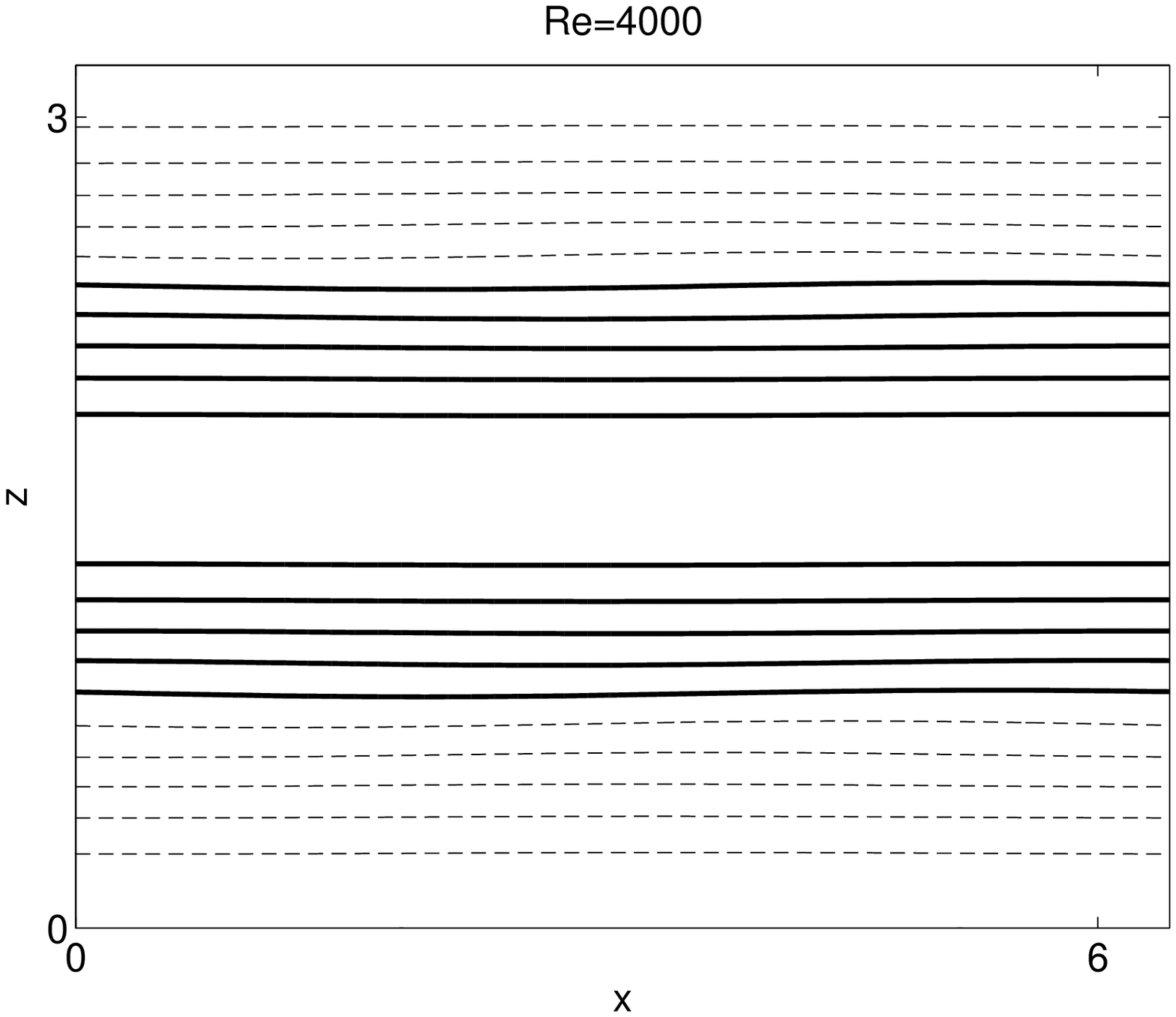}
\end{center}
\caption[xyz]{Contour plots of the streamwise velocity at $y=0$.
The plots correspond to $D1$, $C1$, and $C4$. Contour lines are drawn
at $12$ equispaced values between the maximum and minimum streamwise
velocity in the slice. The lines are solid for positive values and
dashed for negative values. The minimums are $-0.1922, -0.3969,
-0.3833$ and the maximums are $0.4146, 0.3969, 0.3833$. In each plot
the maximum occurs in the widest gap between the solid lines.}
\label{fig-5}
\end{figure}

A notable feature of the solutions of Tables \ref{table-1}, \ref{table-2}
and \ref{table-3} is that the solutions are streaky. This feature
is illustrated in Figure \ref{fig-5}. The contour lines for the streamwise
velocity are approximately parallel to the $x$ axis, but the streamwise
velocity varies in a pronounced way in the $z$ direction. We observe that
$D1$ is less streaky than $C1$. The contour lines become much
straighter when we go from $C1$ to $C4$. This increase in streakiness
with $Re$ is in accord with the asymptotic theory sketched in \cite{WGW}.

To show that these solutions are not fully turbulent, we begin by
describing the use of frictional or wall units \cite{MY}. The mean
shear at the wall, which is denoted by $\bigl<\frac{\partial
u}{\partial y}\bigl|_{y=1}\bigr>$, is the basis for frictional
units. The frictional units for velocity and length are given by
\begin{equation*}
u_f = \sqrt{\nu \Bigl<\frac{\partial u}{\partial
y}\Bigl|_{y=1}\Bigr>}
\quad \quad \text{and} \quad\quad 
l_f = \nu/u_f,
\end{equation*}
respectively. If the width of the channel is $L$, the frictional
Reynolds number is given by $Re_\tau = L u_f/\nu = L/l_f$.
The width of the channel in frictional units equals the frictional
Reynolds number. The use of frictional units is signaled by using
$+$ as a superscript.

The use of frictional units is necessary to state some remarkable
properties of turbulent boundary layers. If $y^+$ measures the
distance from the wall and $<\!\!u\!\!>^+$ is the mean streamwise
velocity in frictional units, after making $<\!\!u\!\!>^+=0$ at
$y^+=0$ by shifting the mean velocities if necessary, then
$<\!\!u\!\!>^+\approx y^+$ in the viscous sublayer.  The viscous
sublayer is about $5$ frictional units thick.  The buffer layer
extends from $5$ to about $30$ units. It is followed by the
logarithmic layer where $<\!\!u\!\!>^+
\approx A \log y^+ + B$, for constants $A$ and $B$. These
relationships between $<\!\!u\!\!>^+$ and $y^+$ have been confirmed in
numerous experiments and in some computations. The experiments are of
a very diverse nature as discussed in \cite{MY}, and it is remarkable
that such a simple relationship holds across all those experiments.

There are other relationships that govern the dependence of quantities
such as turbulence intensities or turbulent energy production on the
distance from the wall. These relationships also characterize
turbulent boundary layers. To show that the $C$ and $D$ solutions are
not fully turbulent, we will use plots of turbulent energy production.
Turbulent energy production equals $$ - <\!\!u^\ast v^\ast\!\!>
\frac{\partial <\!\!u\!\!>}{\partial y},$$ where $u^\ast = u -
<\!\!u\!\!>$ and $v^\ast = v - <\!\!v\!\!>$ are the fluctuating
components of the streamwise and wall-normal velocities and
$<\!\!u\!\!>$ is the mean streamwise velocity.  Turbulent energy
production is easy to measure experimentally and shows a very sharp
peak in the buffer region of turbulent boundary layers
\cite{KRSR}. This sharp peak has intrigued experimentalists for a long
time. In experiments, the means are calculated by averaging pointwise
measurements over long intervals of time.  The means involved in the
definition of turbulent energy production will be computed by
averaging in the $x$ and $z$ directions.

\begin{figure}
\begin{center}
\includegraphics[height=2.1in, width=2.1in]{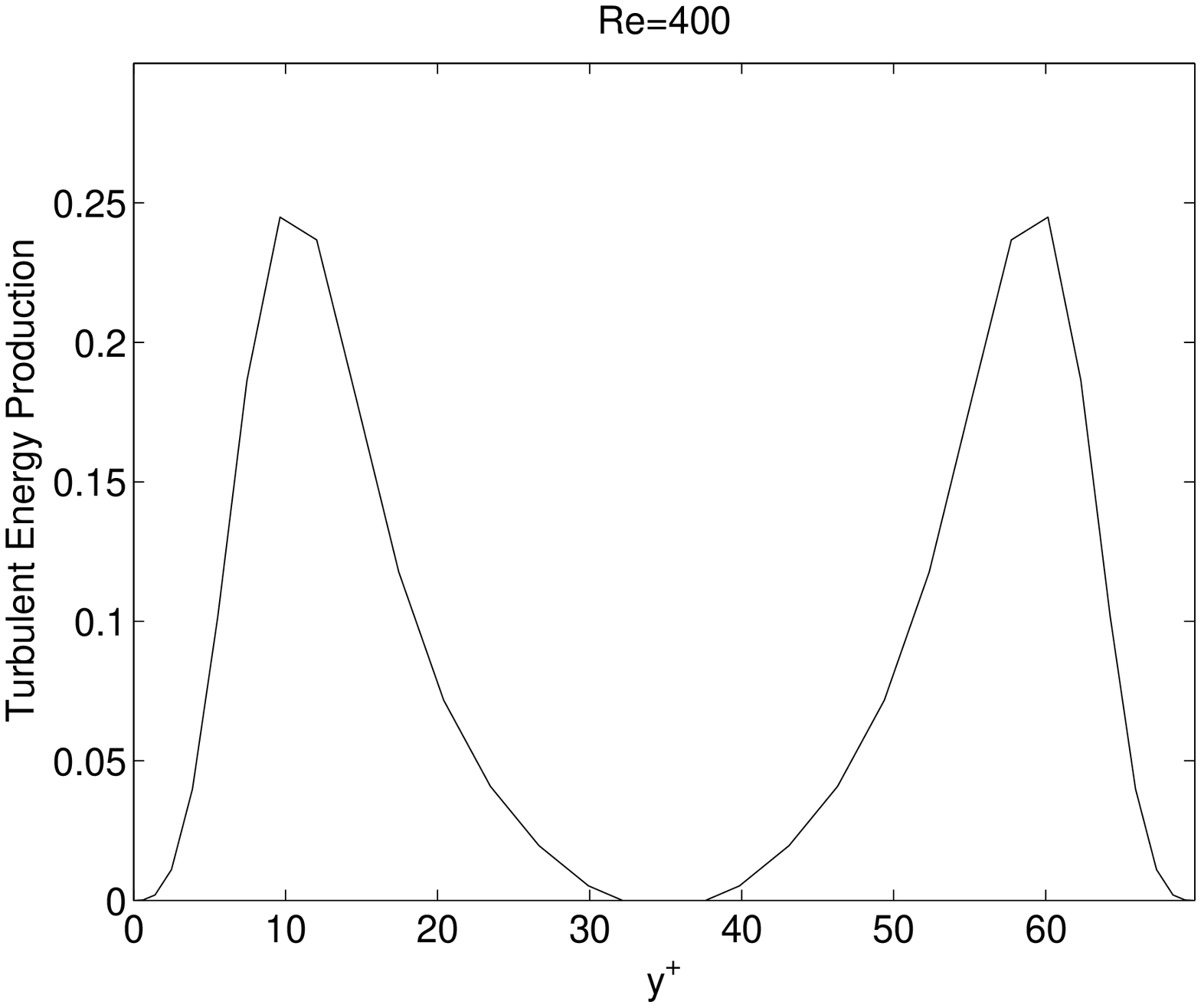}
\includegraphics[height=2.1in, width=2.1in]{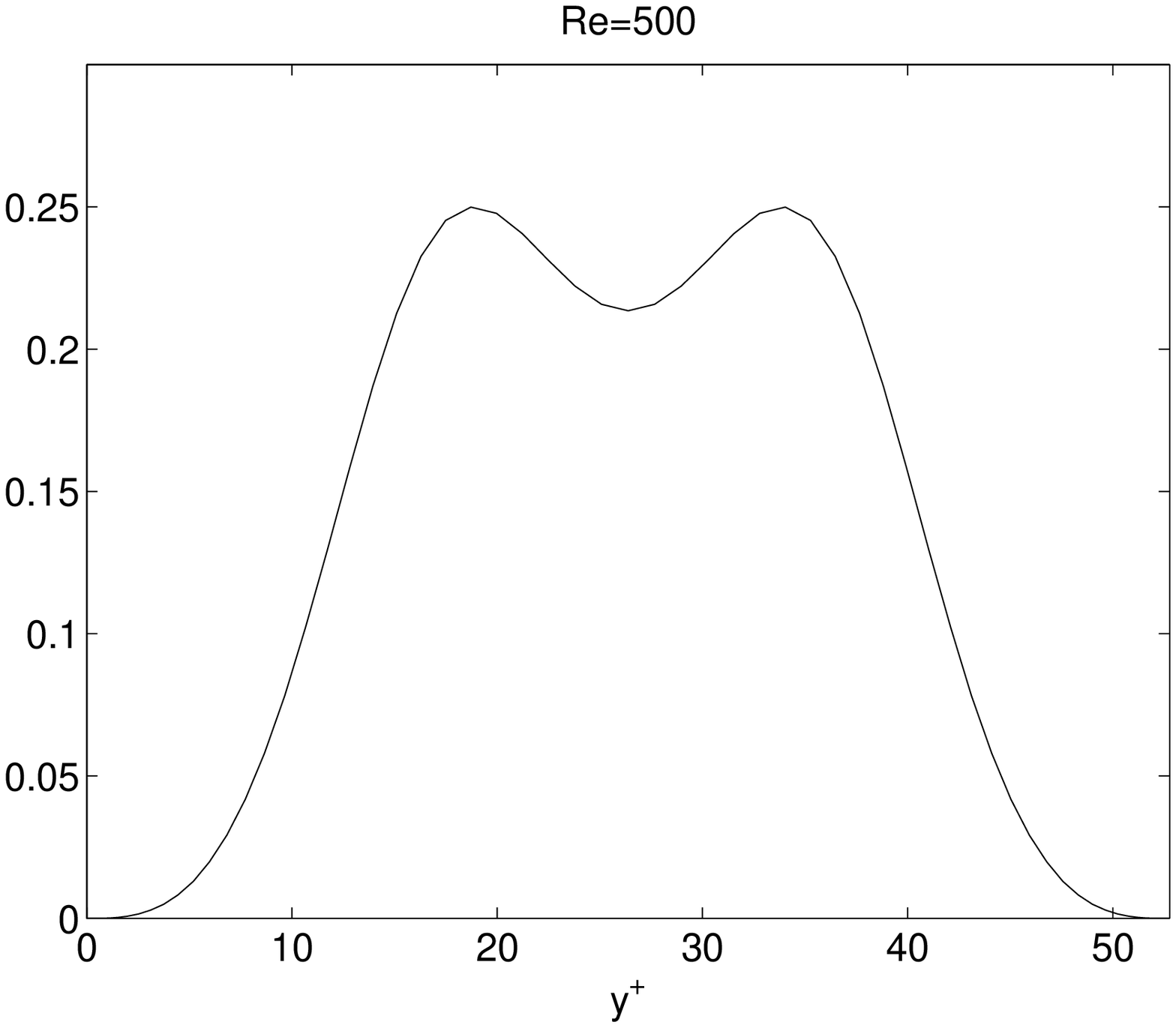}
\includegraphics[height=2.1in, width=2.1in]{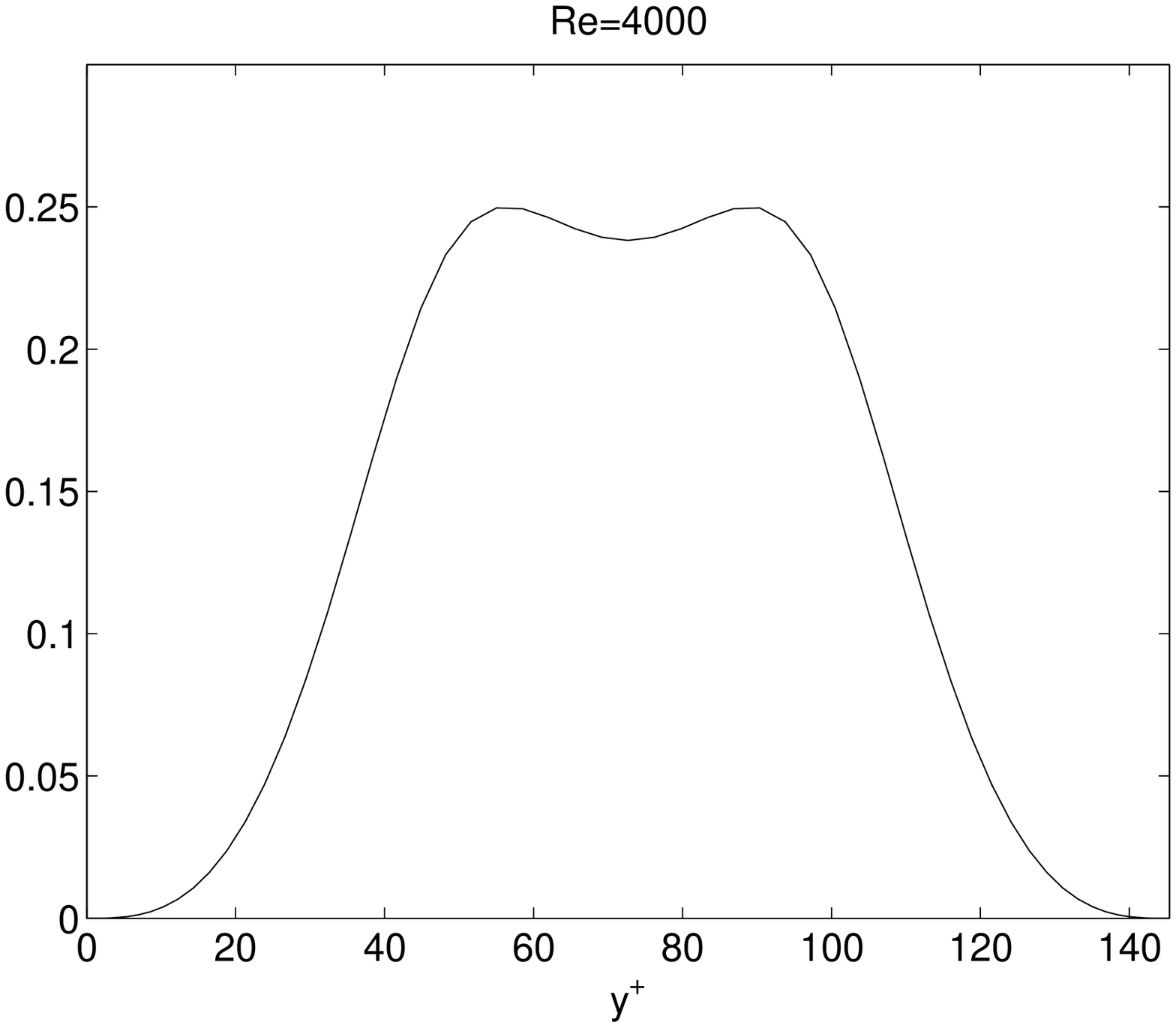}
\end{center}
\caption[xyz]{The plots show the dependence of turbulent 
energy production in frictional units on $y^+$ for
a turbulent steady solution, $C1$, and $C4$.}
\label{fig-6}
\end{figure}

Figure \ref{fig-6} shows plots of turbulent energy production against
$y^+$, the distance from the upper wall in frictional units. In each
plot, $y^+$ varies from $0$ to the channel width. The first plot is
for a turbulent steady solution of plane Couette flow at $Re=400$.
The data for the velocity field of that solution is from
\cite{Waleffe3}.  The second and third plots are for $C1$ and $C4$,
respectively. The first plot is strikingly different from the other
two. In the first plot, we notice that turbulent energy production
peaks inside the buffer layer and then falls off sharply, in a way
that is typical of turbulent boundary layers. The second and third
plots correspond to higher $Re$, yet the peak occurs farther away from
the wall in frictional units and there is no sharp fall-off.  The
plots for $D1$ and $D2$ are not shown.  Those plots are similar to the
ones for $C1$ and $C4$ in that they do not match what we expect for
turbulent boundary layers. A notable difference is that the plots for
$D1$ and $D2$ are not symmetric about the center of the channel. Thus
the $C$ and $D$ solutions exhibit some aspects of near-wall turbulence
such as the formation of streaks, but do not exhibit many other
aspects.

\begin{figure}
\begin{center}
\includegraphics[height=2.1in, width=2.1in]{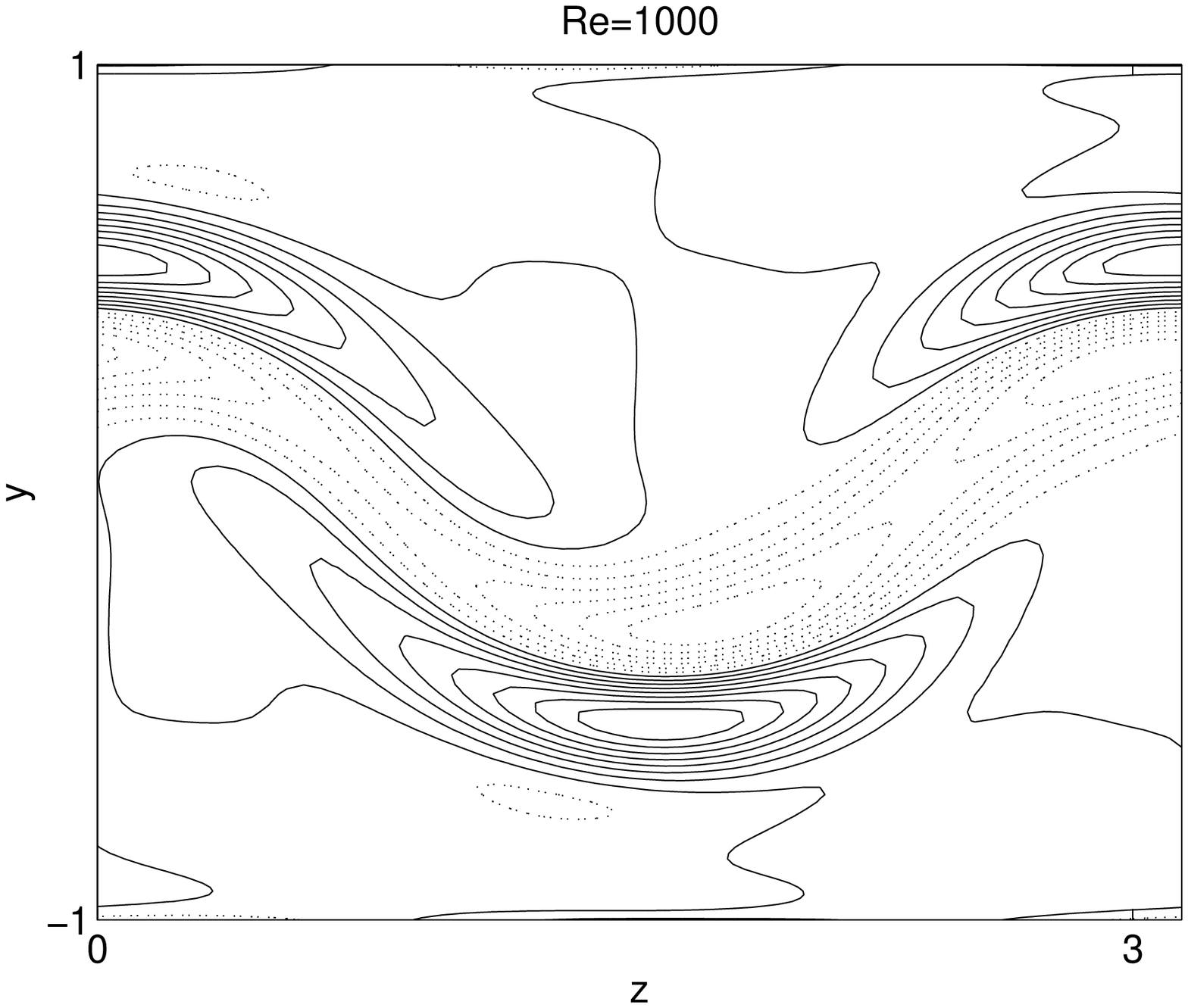}
\includegraphics[height=2.1in, width=2.1in]{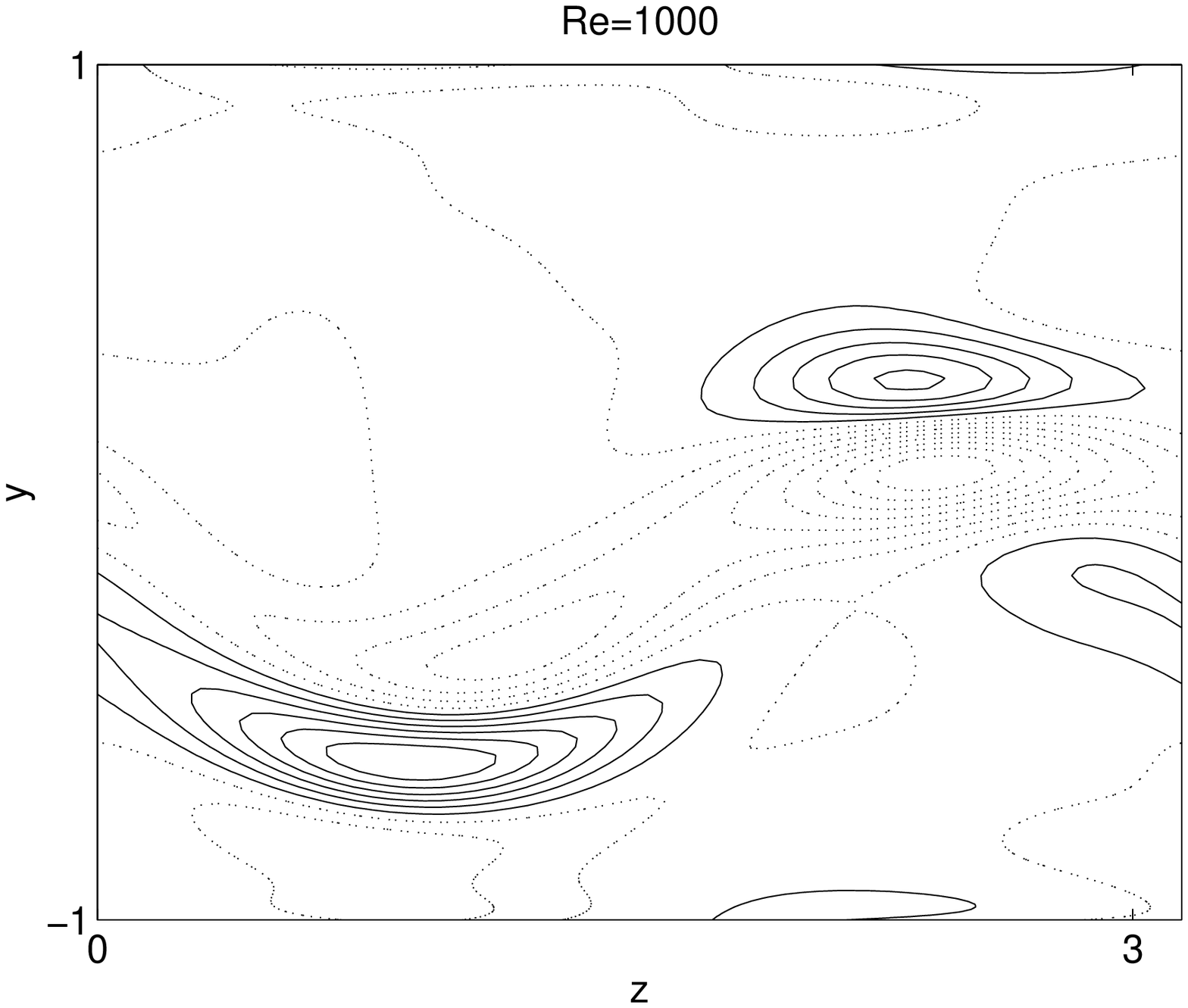}
\end{center}
\caption[xyz]{Contour plots of the streamwise vorticity at
$x=\pi$. The contour lines are equispaced between $-0.11$ and $0.13$
for the first plot, which corresponds to $C2$, and between
$-0.19$ and $0.17$ for the second plot, which corresponds to
$D2$. The lines are dotted for negative values of streamwise vorticity.}
\label{fig-7}
\end{figure}

Figure \ref{fig-7} is another illustration of the qualitative similarity
between the $C$ and $D$ solutions. In both plots of Figure \ref{fig-7},
one may observe a region near the center of the channel where
the streamwise vorticity varies rapidly. Those regions correspond
to the critical layer discussed in \cite{WGW}.

\section{Conclusion}

We verified the dynamical picture for transition to turbulence given
in Figure \ref{fig-1} for certain disturbances. The
third type of disturbance considered in Section 4.3 shows that
that picture does not hold for all disturbances. A more exhaustive study
of different types of disturbances of the laminar solution would
be desirable.

We found (along with Wang et al.\!\! \cite{WGW})
that the $B$ or $C$ solutions become less unstable as
$Re$ increases. This was an unexpected finding. Even a good
heuristic explanation of this trend would be interesting.

Transition to turbulence computations would be good
targets for reduced dimension methods. Reduced dimension methods are
diverse in nature. Although this is not the place to review them, we
believe the intricate dynamics of transition of turbulence featuring
steady solutions, traveling waves, thresholds and various types of
disturbances makes it non-trivial to reduce dimension.  A valid way to
reduce dimension must capture the dynamics correctly and not introduce
spurious artifacts. It has been known since the work of Orszag and
Kells \cite{OK} that under-resolved spatial discretizations lead to
spurious transitions.

It is important to connect transition computations to experiments.
However, connecting transition computations to experiments is impeded
by two problems. Firstly, the experiments are performed in much larger
domains to eliminate boundary effects. The numerical methods reviewed
and discussed in Section 2 ought to be able to handle at least $10$
million degrees of freedom with a good parallel implementation.
Therefore it seems that computations can be performed in much larger
domains (\ie, domains with larger $\Lambda_x$ and $\Lambda_z$) and
that this problem can be overcome. Secondly, it is very difficult to
imagine a way to reproduce the sort of disturbances that have been
considered in the computational literature in experiments. The
disturbances used in experiments are of a different sort.  For
instance, one type of disturbance is to inject fluid from the
walls. The best way to reconcile this disparity between computation
and experiment might be to carry out computations using good models of
laboratory disturbances.

\section{Acknowledgments}
 
The author thanks B. Eckhardt, 
J.F. Gibson, N. Lebovitz, L.N. Trefethen, and
F. Waleffe for helpful discussions.  This work was supported by the
NSF grant DMS-0407110 and by a research fellowship from the Sloan
Foundation.

\bibliography{references}
\bibliographystyle{plain}
\end{document}